\documentclass[aip,apl,preprint]{revtex4-1}
\usepackage{graphicx}
\usepackage{dcolumn}
\usepackage{bm}
\usepackage{color,graphicx}
\usepackage{booktabs}
\usepackage{rotating}

\begin{document}

\title{Spin filtering in nanowire directional coupler}

\author{M. Rebello Sousa Dias}
\affiliation{Departamento de F\'{i}sica, Universidade Federal de S\~{a}o Carlos, 13565-905, S\~{a}o Carlos, S\~{a}o Paulo, Brazil}
\affiliation{Department of Physics and Astronomy and Nanoscale and Quantum Phenomena Institute, Ohio University, Athens, Ohio 45701-2979}
\affiliation{Dahlem Center for Complex Quantum Systems and Fachbereich Physik, Freie Universit\"{a}t Berlin, 14195 Berlin, Germany}
\author{V. Lopez-Richard}
\affiliation{Departamento de F\'{i}sica, Universidade Federal de S\~{a}o Carlos, 13565-905, S\~{a}o Carlos, S\~{a}o Paulo, Brazil}
\author{G. E. Marques}
\affiliation{Departamento de F\'{i}sica, Universidade Federal de S\~{a}o Carlos, 13565-905, S\~{a}o Carlos, S\~{a}o Paulo, Brazil}
\author{S. E. Ulloa}
\affiliation{Department of Physics and Astronomy and Nanoscale and Quantum Phenomena Institute, Ohio University, Athens, Ohio 45701-2979}
\affiliation{Dahlem Center for Complex Quantum Systems and Fachbereich Physik, Freie Universit\"{a}t Berlin, 14195 Berlin, Germany}

\begin{abstract}

The spin transport characteristics of a nanowire directional electronic coupler have been evaluated theoretically via a transfer matrix approach. The
application of a gate field in the region of mixing allows for control of spin current through the different leads of
the coupler via the Rashba spin-orbit interaction.
The combination of spin-orbit interaction and applied gate voltages on different legs of the coupler give rise to
a controllable modulation of the spin polarization.  Both structural factors and field
strength tuning lead to a rich phenomenology that could be exploited in spintronic devices.

\end{abstract}

\maketitle

The nature of electron and spin transport mechanisms can be unveiled by exploring the properties of coupled nanowires (NW). The
realization of a directional coupler gave rise, through proximity and tunneling effects,\cite{couplers, couplersexp} to the modulation of
quantum transport in a phase-coherent system.\cite{reggiani} The spin modulation in a single NW via spin-orbit interaction (SOI) has been
proposed~\cite{dattadas} and refined~\cite{auslaender,auslaender2,naturephys,Loss,Xu} to achieve a spin-orbit quantum bit
device.~\cite{nature} More recently, SOI in semiconductor NWs has enabled the possible observation and characterization of Majorana
fermions.~\cite{Oppen, dassarma, majoranaexp, majoranaexp2} Exploring all-electrical spin transport has motivated a search for new
configurations of nanostructures~\cite{philippe,molenkamp,Ritchie,Xu2} and a wave guide directional coupler
appears as a promising device geometry.

\begin{figure}[hbt]
\linespread{1.0}
\begin{center}
\includegraphics[scale=1]{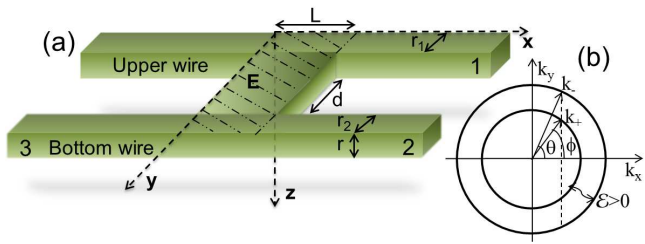}
\end{center}
\vspace{-0.65cm}\caption{(a) Scheme of nanowire (NW) directional coupler with two parallel NW's connected in the middle region by the same
material. The $z$-thickness of the entire region is $r$, $r_{1}$ is the $y$-width of the upper wire, $r_{2}$ of the bottom wire, $d$ is
the NW separation, $L$ is the length of the mixing region, and $\mathbf{E}$ is the electric field along the $z$-direction generated by the
local gate in the shaded central region. (b) Spectrum of 2DEG with Rashba SOI at fixed energy $\mathcal{E}$. For propagation with positive
fixed $x$-component one has four possible values of $k_{y}$ ($\pm k_{-}\sin\theta$, and $\pm k_{+}\sin\phi$), as indicated.} \label{wire}
\end{figure}

In this work, we study the spin transport properties of parallel NWs, in a directional ``H-shape"
coupler geometry, connected through a region of the
same material but locally gated, Fig.~\ref{wire}(a). The application of a gate field ($\mathbf{E}$) in the connecting
region generates a
Rashba SOI, which breaks spin degeneracy,~\cite{rashba} Fig.~\ref{wire}(b).  While the Rashba field does not break time
reversal symmetry,
the symmetries of the configuration impose additional constraints on the transport characteristics.~\cite{Xu, Xu2, Xu3}

We find that, while being able to control the electronic flux through the different wires across the mixing region,
as expected, it is
also possible to electrically control the {\em spin flux} across the device, whenever spin-polarized injection is considered. Moreover, we find
that there is a net spin-polarized flux perpendicular to the current, which can be controlled by the gate potential in the mixing region,
as well as by local gates at the NWs. This overall control of charge and spin flux in a directional coupler appears promising for
spintronics, as well as in hybrid devices composite with superconducting or magnetic materials.

The transfer matrix method\cite{tsu} was used to calculate the transport characteristics of the system. The single-particle
Hamiltonian with $xy$-plane dynamics has the form
\begin{equation}\label{hamiltonian}
\hat{H} =\frac{\mathbf{\hat{p}}^{2}}{2m
}+V_{c}(y,z)+\frac{1}{2\hbar}[\alpha(\mathbf{\mathbf{\mathbf{r}}})\cdot(\sigma_{x}\hat{p}_{y}-\sigma_{y}\hat{p}_{x})],
\end{equation}
where $\hat{p}_{i}=-i\hbar\nabla_{i}$, $\sigma_{i}$ are the Pauli matrices, $\alpha(\mathbf{\mathbf{r}})$ is the Rashba SOI interaction
(proportional to the gate field $\mathbf{E}$), and $V_{c}(y,z)$ is the hard wall confinement potential in the $y$- and $z$-direction. The
eigenfunctions for the Hamiltonian (\ref{hamiltonian}), for vanishing $\alpha(\mathbf{\mathbf{r}})$, in the NWs with square cross section
are given by
\begin{equation}\label{wf}
|\phi_{j}^{\sigma}\rangle=\sqrt{\frac{2}{r_{1,2}}}\sin\left(\frac{n_{y}\pi
y}{r_{1,2}}\right)\sqrt{\frac{2}{r}}\sin\left(\frac{n_{z}\pi z}{r}\right)e^{ik_{x}x}|\sigma\rangle,
\end{equation}
with corresponding eigenvalues
\begin{equation}\label{energ}
\mathcal{E}^{0}_{k_{x},n_{y},n_{z},\sigma}=\frac{\hbar^{2}}{2m}\left(k_{x}^{2}+\frac{\pi^{2}n_{y}^{2}}{r_{1,2}^{2}}+\frac{\pi^{2}n_{z}^{2}}{r^{2}}\right).
\end{equation}
The index $j$ refers to each wire branch: $j=1$ corresponds to the upper wire on the right side; $j=2$ to the bottom right wire; and $j=3$
to the bottom left wire (see Fig.~\ref{wire}(a)). Also, $|\sigma\rangle$ indicates the spinor, and $n_{y}$ and $n_{z}$ are quantum numbers
for the lateral and vertical confinement, respectively. For simplicity, we consider $n_{z}=1$ and two transport channels, $n_{y}=1,2$.

Treating the intermediate mixing region with SOI perturbatively, where $\alpha(\mathbf{\mathbf{r}})\neq0$, the first order correction is dominated by the intrasubband terms ($n_y=n_y'$),  $\langle
\phi'|\alpha\hat{H}_{1}|\phi\rangle \simeq -\frac{i}{2} \alpha k_x \sigma \delta_{\sigma',-\sigma}$,
where, $\sigma, \sigma'=\pm1$, for spin up and down, respectively.
This gives rise to the eigenvalue equation,
\begin{eqnarray}
\left( \begin{array}{cc}
0 & \frac{i\alpha k_{x}}{2} \\
\frac{-i\alpha k_{x}}{2} & 0 \end{array} \right)\left(\begin{array}{c}
C^{\uparrow}  \\
C^{\downarrow} \end{array}\right)= \mathcal{E}^{(1)} \left(
\begin{array}{c}
C^{\uparrow}\\
C^{\downarrow} \end{array} \right),
\end{eqnarray}
which gives the first order correction to the energy levels $ \mathcal{E}^{(1)}_{\pm}=\pm\frac{\alpha k_{x}}{2}$,
and  respective eigenvectors
\begin{equation}\label{mf}
|\psi_{n}^{\sigma}\rangle_{\pm}=\frac{1}{\sqrt{2}}\left[|\phi_{n}^{\sigma}\rangle\pm i|\phi_{n}^{-\sigma}\rangle\right].
\end{equation}
Hence, the wave number component in the $x$-direction for a given energy $\mathcal{E}$ can be written
in the mixing region as
\begin{equation}\label{kxm}
k_{n}^{\pm}=\sqrt{\frac{2 m \mathcal{E}}{\hbar^{2}}-\frac{\pi ^{2} n^{2}}{(d+r_{1}+r_{2})^2}-\frac{\pi^{2}}{r^{2}}+\frac{\alpha ^{2}
m^{2}}{4 \hbar^{4}}}\mp\frac{\alpha m}{2 \hbar^{2}}.
\end{equation}

For the transfer matrix calculation the system is divided into three regions: the regions where $x<0$ (left)
and $x>L$ (right) with no SOI ($\alpha=0$), and
$0<x<L$ (middle region) with SOI ($\alpha\neq0$), as depicted in Fig.\ \ref{wire}. The interface between these regions is considered
sharp for simplicity, and the boundary conditions result in the relations~\cite{boundary}
\begin{eqnarray}\label{boundaryrelation}
\frac{\partial\Psi(\epsilon_{0})}{\partial x}+\frac{m i\sigma_{y}}{\hbar^{2}}\alpha(\epsilon_{0})\Psi(\epsilon_{0})=\\ \nonumber
\frac{\partial\Psi(-\epsilon_{0})}{\partial x}+\frac{m i\sigma_{y}}{\hbar^{2}}\alpha(-\epsilon_{0})\Psi(-\epsilon_{0}),
\end{eqnarray}
where the mass is assumed to be the same everywhere, and  $\epsilon_{0}\rightarrow0$ or $L$ at the two interfaces.
An $8\times8$ transfer matrix is built relating the
coefficients (incident, transmitted, and reflected waves for spin up and down) for a given incident spin,
for the bottom and upper wires
\begin{eqnarray}\label{buildingTMM}
\left( \begin{array}{c}
\mathbf{I}  \\
\mathbf{R}\end{array} \right)=\mathbf{M}_{{\rm Left}}(0)^{-1}\mathbf{M}_{{\rm Middle}}(0) \times\\
\nonumber \mathbf{M}_{{\rm Middle}}(L)^{-1}\mathbf{M}_{{\rm Right}}(L)\left(
\begin{array}{c}
\mathbf{T}  \\
\mathbf{I'}
\end{array} \right),
\end{eqnarray}
where $\mathbf{T}=T_{j}^{\sigma'\sigma}$, $\mathbf{R}=R_{j}^{\sigma'\sigma}$, and $\mathbf{I'}=I_{j}^{'\sigma'\sigma}$ are all four
dimensional vectors for the transmitted, reflected, and incident amplitudes for the different NW branches, with incident spin $\sigma'$,
and outgoing spin $\sigma$.
The overlap terms, which contain information on the geometry and lateral confinement are included in the matrices
$\mathbf{M}_{{\rm Middle}}(0)$ and $\mathbf{M}_{{\rm Middle}}(L)$.

The quantum transport properties of interconnected parallel wires
through a potential barrier \cite{couplers, reggiani, zulicke, Lilly} can be controlled by changing the barrier height and
length.  Similarly, one can modulate the carrier and spin transport by varying the length $L$ of the middle mixing region
and the separation $d$ between wires.
The formalism enables the study of different spin polarization configurations for the incident waves.
We present results for an incident wave in one
of the wires with up or down spin-polarization along the $z$-direction.
The transmission and reflection amplitudes (the entire scattering matrix) will be shown below to exhibit oscillations
arising from the expected interference among the various channels, as they mix in the coupling region.
This interference has
characteristic energy (or length) scales associated with the different components producing the interference.
We will also see that the onset of spin-orbit coupling in
the mixing region further complicates the pattern of oscillations, as SOI effectively duplicates the number of available channels at a given energy.

\begin{figure}[hbt]
\linespread{1.0}
\begin{center}
\includegraphics[scale=1]{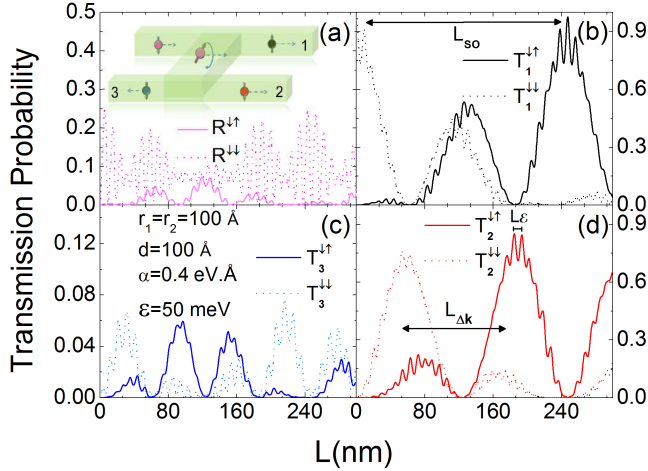}
\end{center}
\vspace{-0.65cm}\caption{Transmission probabilities vs length of the mixing region $L$ for incident spin-down polarized flux
($I^{\downarrow}$), where $r_{1}=r_{2}=100$ \AA, $d=100$ \AA, $\mathcal{E}=50$ meV, $\alpha=0.4$ eV$\cdot$\AA, and $m=0.026 m_{0}$. (a)
Reflection on the left top wire ($R^{\downarrow\sigma}$). (b) Transmission on the right top wire ($T_{1}^{\downarrow\sigma}$). (c)
Transmission on the left bottom wire ($T_{3}^{\downarrow\sigma}$). (d) Transmission on the right bottom wire ($T_{2}^{\downarrow\sigma}$).
Notice reflection values in (a) and (c) are much smaller than the transmission in (b) and (d).}\label{LSOI}
\end{figure}

Figure \ref{LSOI} shows the transmission and reflection coefficients vs the length of mixing region $L$ for an incident spin-down polarized
wave (inset Fig.~\ref{LSOI}), where $d=100$ \AA, and $r_{1}=r_{2}=100$ \AA. We can see in Fig.\ \ref{LSOI}(d) that the spin-preserving transport
coefficient $T_{2}^{\downarrow\downarrow}$ can be suppressed or enhanced with a characteristic
periodicity  $\simeq L_{\Delta k}=2 \pi/\Delta k $.
For a given incident energy $\mathcal{E}$, there are corresponding
wave vectors $k^{\pm}_{n}$ along each given channel, $n=1,2$ (Eq.~(\ref{kxm})). The momentum difference $\Delta
k=k_{1}^{\sigma}-k_{2}^{\sigma}$, is responsible for the long period oscillations as the two ($\pm$) channels
interfere with one another.
For $\mathcal{E}=50$ meV, one finds
$L_{\Delta k}=2\pi/\Delta k \simeq 120$ nm, a value that coincides with the oscillation period.
Moreover, whenever $T_{2}^{\sigma'\sigma}$ is enhanced,
$T_{1}^{\sigma'\sigma}$ is suppressed (Fig.\ \ref{LSOI}(b) and (d))--i.e,
they oscillate out of phase, illustrating their competition.
Opposite phase oscillations between the
NWs result for a symmetric system such as this, where the initial conditions have the incident wave in only
one of the coupled NW's (inset Fig.\ \ref{LSOI}).
The coefficients also show fast oscillations characterized by $L_{\mathcal{E}}=\sqrt{\pi^{2}\hbar^{2}/2
m\mathcal{E}}$, in analogy with the quantum well system. For $\mathcal{E}=50$ meV, these fast oscillations
have the corresponding length scale $L_{\mathcal{E}}\simeq 10$ nm. Thus, one is able to adjust the transmission
by changing sizes, widths, and lengths of mixing
regions, as expected from the directional coupler geometry of the system.\cite{couplers, couplersexp, zulicke}

The introduction of SOI provides additional control on the overall transmission amplitudes, as well as on
the spin polarized transmission.
The SOI length scale, given by the period over which the spin precesses from $\uparrow$ to $\downarrow$
(and vice versa) due to the effective Rashba field, is $L_{SO}=\pi\hbar^{2}/\alpha m$
($\simeq 230$ nm, in Fig.\ \ref{LSOI}). One can clearly see the effect of this precession in
Fig.\ \ref{LSOI}(b), where the incident spin down
is fully transmitted in the up-projection at $L \simeq L_{SO}$, $T^{\downarrow \uparrow}_1$, while the
spin-preserving channel is nearly
fully blocked, $T^{\downarrow \downarrow}_1 \simeq 0$.
Figure \ref{LSOI}(a) and (c) show the reflection ($R^{\downarrow\sigma}$) and
transmission in the bottom wire on the left NW ($T_{3}^{\downarrow\sigma}$), respectively.
These coefficients are characterized by
oscillations with shorter periods due to the fact that the wave travels twice the distance $L$.
Moreover, in this configuration the length
scale $L_{SO}$ becomes comparable to $2L_{\Delta k}$, resulting in a more complex interference and the
beating behavior evident on
$R^{\downarrow\downarrow}$. These results are for a spin-polarized incident wave; reversing the incident spin
produces the spin reversal of the transmitted and reflected amplitudes, as expected from the time-reversal
invariance of the Hamiltonian, as well as
from the spatial symmetries of the system.\cite{Xu, Xu2}

\begin{figure}[hbt]
\linespread{1.0}
\begin{center}
\includegraphics[scale=1]{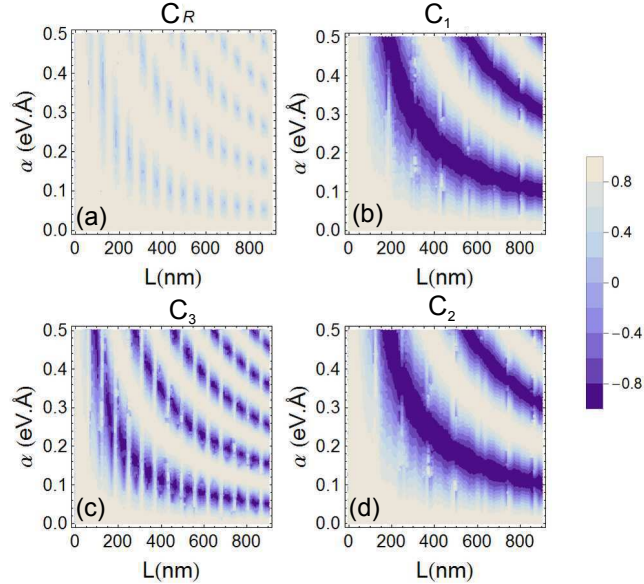}
\end{center}
\vspace{-0.65cm}\caption{Map of spin persistence ratio $C_{j}$ vs $\alpha$ and $L$, where $r_{1}=r_{2}=100$ \AA, $d=100$ \AA, and
$\mathcal{E}=50$ meV. Panels show results for different NWs: (a) top left reflection, (b) top right, (c) bottom left, and (d) bottom right
transmissions. Spin-reversal regions (dark blue) occur for $\alpha L \simeq L/L_{SO} = {\rm constant}$.}\label{CD}
\end{figure}

To better characterize how the system responds to the injection of a spin unpolarized superposition of
spin-up and spin-down fluxes, we calculate the spin persistence ratio
$C_{j}\equiv\left[T^{\uparrow\uparrow}_{j}+T^{\downarrow\downarrow}_{j}-(T^{\downarrow\uparrow}_{j}+T^{\uparrow\downarrow}_{j})\right]/T$,
where $T=T^{\uparrow\uparrow}_{j}+T^{\downarrow\downarrow}_{j}+T^{\downarrow\uparrow}_{j}+T^{\uparrow\downarrow}_{j}$, for each NW $j$.
Note that if $T^{\uparrow\uparrow}_{j}=T^{\uparrow\downarrow}_{j}$ and $T^{\downarrow\downarrow}_{j}=T^{\downarrow\uparrow}_{j}$ then
$C_{j}=0$, a signature of a ``memoryless'' channel; on the other hand, $C_{j}=1$ $(-1)$ for a full spin-preserving
(or reversing) channel.
Figure \ref{CD} shows the color map of characteristic $C_{j}$ vs $\alpha$ and $L$ for each wire branch, $j$. The spin
reversing regions are identified by the dark (blue) color and evolve with both $\alpha$ and length $L$, such that $\alpha L \simeq L/L_{SO} ={\rm constant}$, as expected from the Rashba precession length.
Alternatively, pale gray regions indicate spin-preserving characteristics.  One can also identify nearly
$\alpha$-independent oscillations,
more visible on Fig.\ \ref{CD}(a) and (c) within this area, associated with $L/L_{\Delta k}={\rm constant}$.

\begin{figure}[hbt]
\linespread{1.0}
\begin{center}
\includegraphics[scale=1]{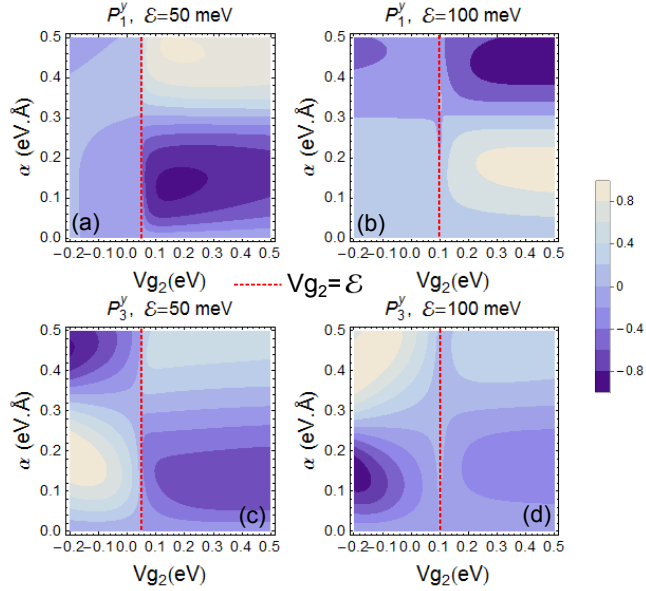}
\end{center}
\vspace{-0.65cm}\caption{Polarization in the $y$-direction vs $\alpha$ and applied gate voltage on the bottom right wire, $(Vg_{2})$,
where $r_{1}=r_{2}=100$ \AA, $d=100$ \AA, and  $L=300$ nm. (a) Polarization for the top right wire ($P^{y}_{1}$) with $\mathcal{E}=50$
meV. (b) Polarization for the top right wire ($P^{y}_{1}$) with $\mathcal{E}=100$ meV. (c) Polarization for the bottom left wire
($P^{y}_{3}$) with $\mathcal{E}=50$ meV. (d) Polarization for the bottom left wire ($P^{y}_{3}$) with $\mathcal{E}=100$ meV.~ Red dashed
lines indicate $Vg_2=\mathcal{E}$ in each panel.}\label{SOIvoltage}
\end{figure}

One cannot expect a net spin polarization for the current, even when the system symmetry is broken by applying a gate voltage difference
between the wires or making the device asymmetric in other ways. This is due to the fact that for transport along the $x$-direction, the
electric field on the $z$-direction, and corresponding Rashba SOI, will produce spin precession of the $z$-spin for both spin
components in a symmetric fashion.~\cite{souma,Xu2} The degree of spin polarization is defined by
$P_{j}\equiv\left[T^{\uparrow\uparrow}_{j}+T^{\downarrow\uparrow}_{j}-(T^{\downarrow\downarrow}_{j}+T^{\uparrow\downarrow}_{j})\right]/T$.
As expected, there is no net polarization in the $z$-direction. However, the effective Rashba magnetic field is
along the $y$-direction, which
suggests one to explore the possibility of tuning this quantity along that axis, $P_j^y$.
The application of a gate voltage in one of
the branches to raise the local energy of the subbands, as well as the strength of the Rashba SOI, can further help to tune the polarization in the remaining output wires.
Figure \ref{SOIvoltage} shows $P_{j}^y$ as a function of the Rashba SOI strength, $\alpha$, and an applied gate
voltage on the bottom right wire, $Vg_{2}$. Panels (a) and (c) correspond to the incident energy $\mathcal{E}=50$ meV, and panels (b) and
(d), to $\mathcal{E}=100$ meV.~ One can see that Fig.\ \ref{SOIvoltage}(a) displays two regions where the $y$-polarization is well defined.
Around $\alpha=0.12$ eV$\cdot$\AA~ and $Vg_{2}=0.14$ eV we notice full spin-down polarization, while for the same $Vg_{2}$ but at $\alpha=0.48
$ eV$\cdot$\AA, one has the opposite polarization. These features change location and sign as the incident energy varies. These results demonstrate
the possibility of spin-inversion of the current flux by controlling $\alpha$ (via $\mathbf{E}$) or $Vg_{2}$ (via applied gates).
Figure \ref{SOIvoltage}(c) and (d) correspond to $T_{3}$, and one may note that they present similar behavior,
but where features have a smoother (not as sudden) variations in $\alpha$ or $Vg_2$.

In conclusion, we have studied the spin transport properties of a nanowire directional coupler implemented by
parallel wires joined by a mixing
region which generates Rashba spin-orbit interaction (SOI). Using a transfer matrix approach allowed us to understand
the modulation of electronic and spin transport arising from the combination of SOI and the system geometrical features.
Likewise, SOI and applied gate voltages give rise
to a modulation of the polarization when the spin is projected in the same direction as the effective Rashba magnetic field.
The versatility of this device may be useful in all-electrical spintronic devices.

The authors acknowledge the support of CAPES, CNPQ, and FAPESP-12/02655-1 (Brazil), the Alexander von Humboldt
Foundation, and DAAD (Germany), as well as NSF grant 1108285 from DMR-MWM/CIAM (USA).

\pagebreak

\end{document}